\documentclass[runningheads]{llncs}
\usepackage{graphbox} %
\usepackage{bm}
\usepackage{amsmath}
\usepackage{todonotes}
\usepackage{booktabs}
\usepackage{graphicx}
\usepackage{mwe}
\usepackage{placeins}

\newcommand{\matr}[1]{\mathbf{#1}}
\newcommand{\vect}[1]{\mathbf{#1}}
\newcommand{\etal}{\textit{et al}.}

\begin{document}
\title{Modelling the Distribution of 3D Brain MRI using a 2D Slice VAE}
\author{Anna Volokitin \and
Ertunc Erdil \and 
Neerav Karani \and
Kerem Can Tezcan \and
Xiaoran Chen \and
Luc Van Gool \and 
Ender Konukoglu}
\authorrunning{A. Volokitin et al.}
\institute{Computer Vision Lab, ETH Z\"urich\\ 
\email{voanna@vision.ee.ethz.ch}}
\maketitle              %
\begin{abstract}
Probabilistic modelling has been an essential tool in medical image analysis, especially for analyzing brain Magnetic Resonance Images (MRI). Recent deep learning techniques for estimating high-dimensional distributions, in particular Variational Autoencoders (VAEs), opened up new avenues for probabilistic modeling. Modelling of volumetric data has remained a challenge, however, because constraints on available computation and training data make it difficult effectively leverage VAEs, which are well-developed for 2D images.  We propose a method to model 3D MR brain volumes distribution by combining a 2D slice VAE with a Gaussian model that captures the relationships between slices.  We do so by estimating the sample mean and covariance in the latent space of the 2D model over the slice direction.  This combined model lets us sample new coherent stacks of latent variables to decode into slices of a volume. We also introduce a novel evaluation method for generated volumes that quantifies how well their segmentations match those of true brain anatomy. We demonstrate that our proposed model is competitive in generating high quality volumes at high resolutions according to both traditional metrics and our proposed evaluation.\footnote{Code is available at \url{https://github.com/voanna/slices-to-3d-brain-vae/}}

\end{abstract}
\section{Introduction}
Generative modeling with Bayesian models have played an important role in medical image computing, yielding very robust systems for segmentation and extracting morphological measurements, especially for brain MRI~\cite{fischl2012freesurfer,Ashburner2005,Leemput2009}. However, the difficulty in using these earlier Bayesian models was the difficulty in defining prior distributions. The challenges in estimating high-dimensional prior distributions forced researchers to use atlas-based systems through non-linear registration, e.g.~\cite{Ashburner2005}, which arguably limited the applications of such models due to the challenges in registration itself. Recently, unsupervised deep learning has yielded powerful algorithms for estimating distributions in high dimensions and opened new avenues for modeling prior distributions for Bayesian models. 
Notably, Variational AutoEncoder models~\cite{KingmaW13} provide access to probability values through the evidence lower-bound, enabling Bayesian approaches to various problems, such as undersampled Magnetic Resonance (MR) image reconstruction~\cite{tezcan2018mr} and outlier detection~\cite{chen2018unsupervised}. Unfortunately, methods leveraging VAEs so far have had to constrain themselves to 2D models or coarser resolution 3D models. 

Training volumetric VAE models remains difficult, due to limitations in available training data and computational resources. Compared to 2D data, 3D data is evidently higher dimensional, posing challenges for estimating probability distributions. 
The number of 3D training examples is relatively low compared to the 2D case. Even large-scale datasets only contain images on the order of thousands. 
Adding to the problem, volumetric VAEs also have a larger number of parameters to be trained and are difficult to fit into memory in GPU systems. 

This means that existing models typically only demonstrate results for downsampled coarse volumetric data.
Works on generating natural videos represented as ``space-time cuboids''~\cite{vondrick2016generating,kratzwald2017improving} have stopped at $3\times64\times64\times32$ size. Kwon~\etal~\cite{kwon2019generation} recently showed high quality generations of brain MR volumes at $64\times64\times64$ image size with their proposed 3D $\alpha$WGAN method, however the method has difficulty scaling to $256\times256\times256$ in our experiments.

To move to 3D data at larger sizes with finer resolution, we propose to instead use (relatively) easy to train 2D variational autoencoders to generate MR image slices.  We can exploit the correlation between  successive slices of an MR volume in a second modelling step that captures the relationship between slices.  By separately encoding all of the slices coming from the same volume using our 2D encoder, over many different volumes,  we can estimate the sample mean and covariance of the latent codes over the slice dimension.  This gives us a model for 3D data and lets us sample from the distribution by generating a new stack of latent codes with the same mean and covariance as the original codes, which, when decoded, correspond to a new consistent MR volume. We show that this simple yet efficient approach yields generated volumes that are competitive with other proposed generation approaches, such as the recently proposed 3D $\alpha$WGAN~\cite{kwon2019generation} at $128\times128\times128$ image size, and outperforms 3D $\alpha$WGAN at $256\times256\times256$ image size on several metrics.

We additionally introduce a novel and  interpretable evaluation measure of the quality of the generated samples.  We segment generated samples using a segmentation network trained on real images and then register generated volumes to real volumes, along with their segmentations. We then compute the Dice's similarity coefficient (DSC)~\cite{dice1945measures} between the registered segmentations, and call this the``Realistic Atlas Score'' (RAS). This procedure allows us to evaluate (a) how well a generated volume can ``pass" as a real volume in the eyes of both a segmentation network and a registration algorithm; as well as (b) how well the anatomy in the generated images match real ones.  Unlike other common evaluation methods for generative models, such as the Inception Score~\cite{salimans2016improved}, the Fr\'echet Inception Distance~\cite{heusel2017gans}, the RAS has a direct anatomical interpretation, which makes it informative for generative modelling of medical images in particular.

\section{Methods}

\subsection{Modeling distribution of 3D images with 2D VAE}
Our model has two components: (1) a variational autoencoder and (2) a sample mean and covariance estimation in the latent space of the encoder.
The encoder maps MR slices to points in an $L$-dimensional latent space $\mathcal{Y}$ and the decoder maps them back to the image space $\mathcal{X}$. We train this model to convergence.

The second part of our model is a collection of $L$ sample mean and covariance estimates over the latent variables in the slice dimension (one covariance estimate for each component of the latent space of the encoder). Using the sample means and covariances, we can sample new sequences of the latent variables that correspond to sequences of slices through an MR-volume.  These samples will have the same sample mean and covariance structure as the original latent codes. The latent variable corresponding to each slice can be decoded individually to an image, and the slices are combined to obtain a complete and consistent MR-volume.  The consistency of the slices is ensured because we compute the mean and covariance the slice direction. 

\begin{figure}[!htb]
    \centering
    \includegraphics[width=1.0\textwidth]{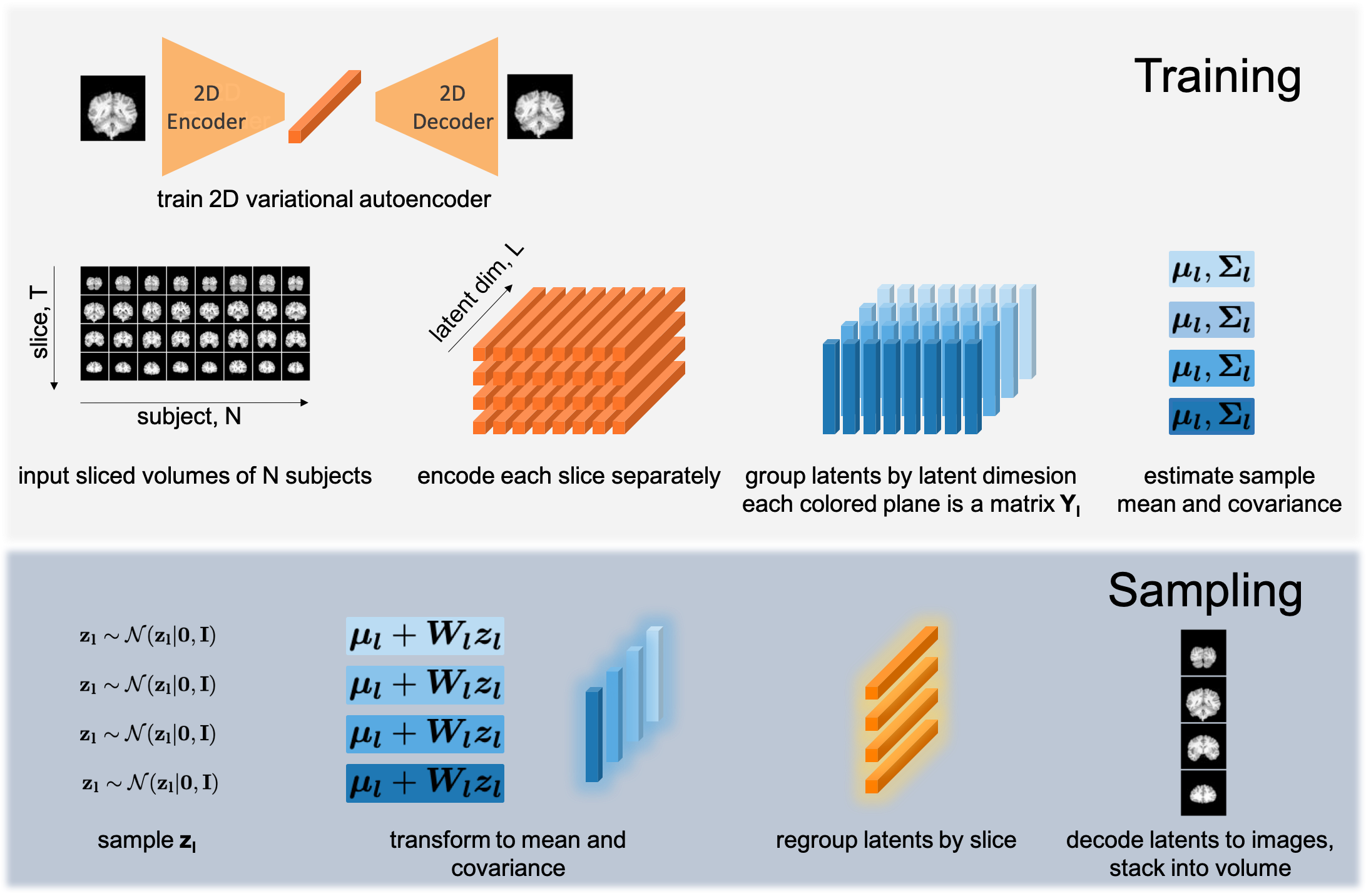}
    \caption{We train a 2D autoencoder model on MR brain slices, and then model the relationship between successive slices in a volume by separately estimating sample means and covariances over the slice dimension for each component of the latent code. Using these, we transform samples from a unit Gaussian into new latent codes that can be decoded into volumes.}
    \label{fig:method}
\end{figure}{}

Specifically, let $\vect{y}(t) = \text{encoder}(\matr{X}(t))$, where $t=1\dots T$ shows the dependence on the slice. Let $y_l(t)$ be the $l$-th component of the latent vector at slice $t$.  We assume that corresponding latent variables across different slices are statistically related and we approximate this relation with a Gaussian model 
\begin{equation*}
    p(\vect{y}_l) = \mathcal{N}(\vect{y}_l| \bm{\mu}_l, \matr{\Sigma}_l),\  \vect{y}_l=\left[y_l(1),\dots, y_l(t), \dots, y_l(T)\right]
\end{equation*}
where $\bm{\mu}_l$ and $\matr{\Sigma}_l$ are the sample mean and covariance matrices at the $l^{th}$ component in the latent space. These sample statistics are computed using the latent representations of the training samples. We encode all a set of training volumes, slice-by-slice, and use the latent codes for estimating the sample statistics.  %

To sample a new $\vect{y}_l$, we can use the expression $\vect{y_l} = \matr{W_l}\vect{z_l} + \bm{\mu_l}$, and sample $\vect{z_l}$ according to $p(\vect{z_l}) = \mathcal{N}(\vect{z_l}|\vect{0}, \matr{I})$
, where $\matr{W}_l = \matr{\Sigma}_l^{1/2}$. To compute $\matr{W}_l$ we use the singular value decomposition of $\matr{Y}_l$, the matrix containing $\vect{y}_l$ for different training samples as columns. If $\matr{Y_l} = \matr{U_l}\matr{S_l}\matr{V_l^*}$, then $\matr{W_l} = \matr{U_l}\matr{S_l}^{1/2}/\sqrt{N}$, where $N$ denotes the number of training samples. For each dimension $l$ in the latent space, we build independent Gaussian models based on sample statistics.

Denoting all the latent variables for a volume together by the vector $\mathbf{y}$, we have $p(\mathbf{y}) = \mathcal{N}(\mathbf{y}|\bm{\mu}\mathbf{_y}, \mathbf{\Sigma_y})$, where 
$\mathbf{y} = \left[\mathbf{y}_1^T, \mathbf{y}_2^T, \dots, \mathbf{y}_L^T\right]^T$, the volume latent mean is
$\bm{\mu_y} = \left[\bm{\mu}_1^T, \dots, \bm{\mu}_L^T\right]^T$, and the volume latent covariance is the block diagonal matrix $
\mathbf{\Sigma_y} = \left[\begin{array}{ccc}
    \mathbf{\Sigma}_1 & \hdots & \mathbf{0}\\
    \vdots & \ddots & \vdots\\
    \mathbf{0} & \hdots & \mathbf{\Sigma}_L
\end{array}\right].\\$

Then decoding each slice of the volume $\mathbf{V}$ individually gives $p(\mathbf{V}|\mathbf{y}) = \prod_t p(\mathbf{V_t} | \mathbf{y}_t)$, where $\mathbf{y}_t = \left[ y_1(t), \dots, y_L(t) \right]$. Together with $p(\mathbf{y})$ from above, the probabilistic model for the entire volume in the proposed approach can be given as $p(\mathbf{V}) = \int p(\mathbf{V}|\mathbf{y}) p(\mathbf{y}) d\mathbf{y}$.

Modelling only slice interactions and assuming independence between latent variables is a simplification that allowed us to have very simple sampling procedure and an explicit form for $p(\mathbf{V})$, as described above.

\subsection{Evaluating quality of the generated samples with RAS}
In addition to the method described above, we propose to use a well-established atlas-based segmentation strategy to evaluate the generated samples by using them as atlases in a segmentation procedure.  This approach is conceptually similar to the Reverse Classification Accuracy (RCA) method~\cite{Valindria2017} that predicts the test-time accuracy of segmentation models. Our procedure is: 
\begin{enumerate}
    \item Segment the generated samples using a CNN-based segmentation network, which is trained using real images
    \item Register the generated samples to real images and map the predicted segmentation with the same transformation. 
    \item Evaluate the agreement between segmentations of the generated samples predicted by the CNN, after mapping, and real images. 
    \item The agreement score between the segmentations serves as the quality metric.
\end{enumerate}

We evaluate the agreement using the DSC and use affine registration. Other choices for agreement metrics and registration algorithms are also possible.

The procedure for computing RAS evaluates the generated samples in three different ways. First, the generated samples has to yield realistic segmentations when fed into the CNN-based segmentation network. To achieve this, they need to be void of any domain-shifts. Second, the generated samples should be ``registerable'' to real images, showing similar intensity profiles across the image. Lastly, the generated samples has to capture correct anatomical details for a high agreement between the segmentations of the generated samples, after mapping, and real images. 

We propose the RAS metric to complement other evaluation scores, such as MMD and MS-SSIM used in~\cite{kwon2019generation}. Previously used scores are aiming to evaluate the diversity of the generated samples more than how realistic they are. RAS aims directly at evaluating realism with a specialized strategy for medical images.

\section{Experimental Setup}
\subsection{Compared models}
\begin{figure}[]
    \centering
    \begin{tabular}{lcccc}
    \toprule
    128$^3$ 3D $\alpha$WGAN & 
        \includegraphics[width=0.2\textwidth,align=c]{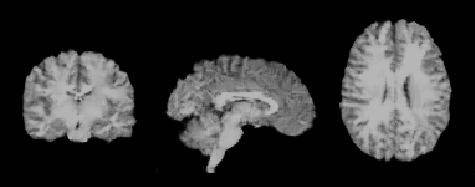}&
        \includegraphics[width=0.2\textwidth,align=c]{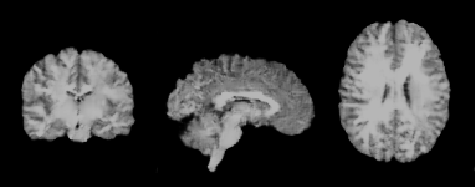}&
        \includegraphics[width=0.2\textwidth,align=c]{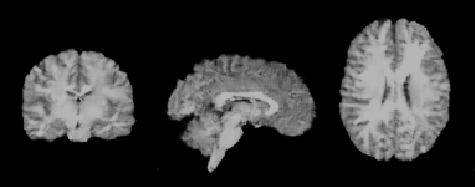}&
        \includegraphics[width=0.2\textwidth,align=c]{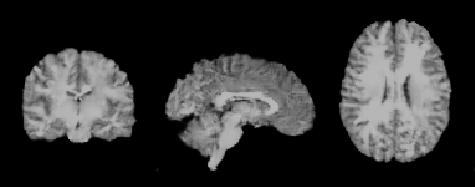}\\
    128$^3$ 3D VAE & 
        \includegraphics[width=0.2\textwidth,align=c]{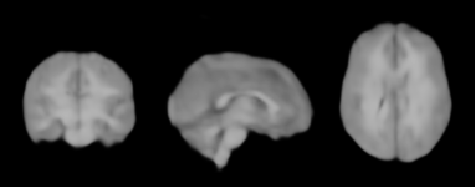}&
        \includegraphics[width=0.2\textwidth,align=c]{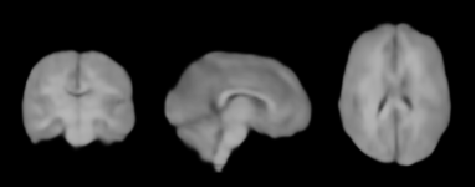}&
        \includegraphics[width=0.2\textwidth,align=c]{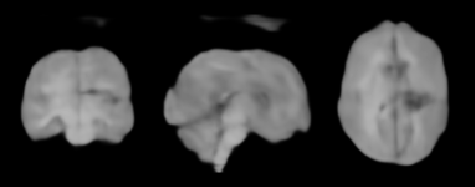}&
        \includegraphics[width=0.2\textwidth,align=c]{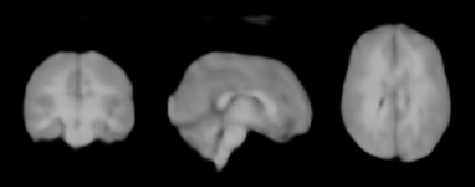}\\
    128$^3$ Ours & 
        \includegraphics[width=0.2\textwidth,align=c]{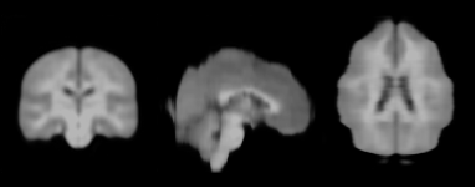}&
        \includegraphics[width=0.2\textwidth,align=c]{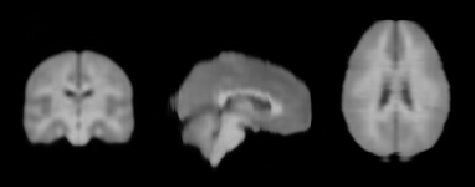}&
        \includegraphics[width=0.2\textwidth,align=c]{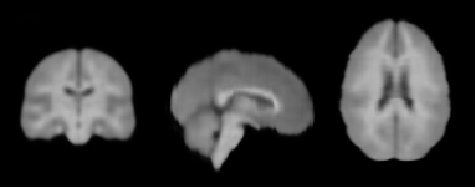}&
        \includegraphics[width=0.2\textwidth,align=c]{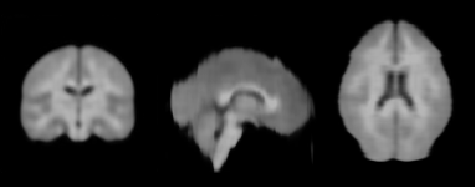}\\
    \midrule
    256$^3$ 3D $\alpha$WGAN & 
        \includegraphics[width=0.2\textwidth,align=c]{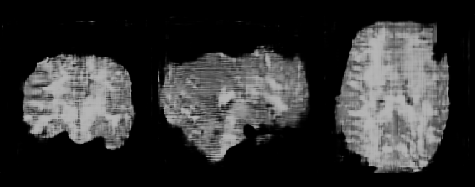}&
        \includegraphics[width=0.2\textwidth,align=c]{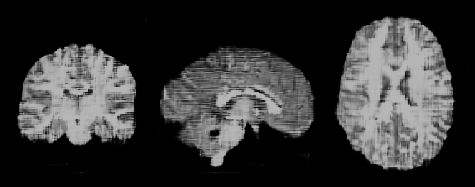}&
        \includegraphics[width=0.2\textwidth,align=c]{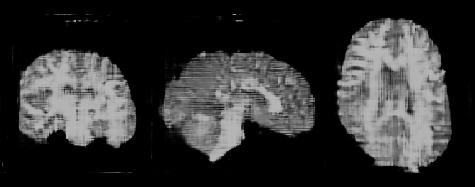}&
        \includegraphics[width=0.2\textwidth,align=c]{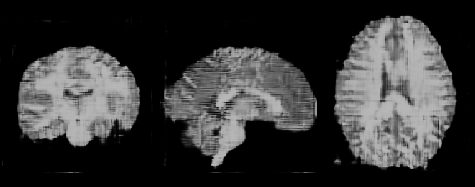}\\
 
    256$^3$ Ours & 
        \includegraphics[width=0.2\textwidth,align=c]{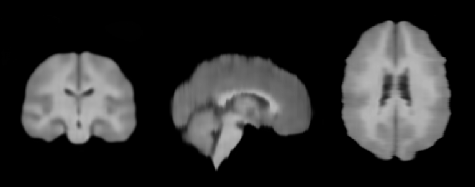}&
        \includegraphics[width=0.2\textwidth,align=c]{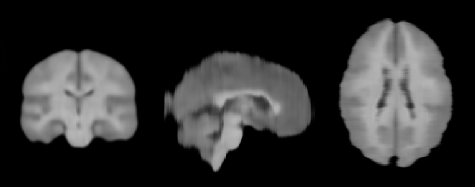}&
        \includegraphics[width=0.2\textwidth,align=c]{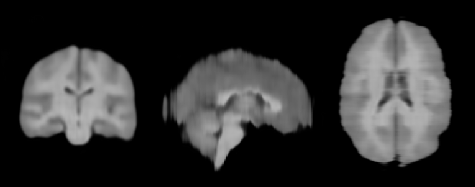}&
        \includegraphics[width=0.2\textwidth,align=c]{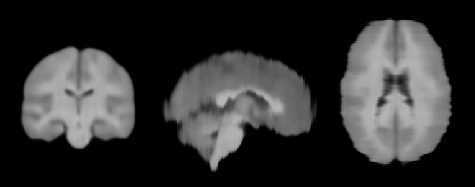}\\
    \midrule
    Real & 
        \includegraphics[width=0.2\textwidth,align=c]{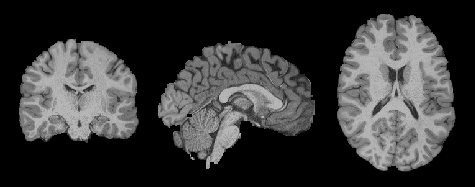}&
        \includegraphics[width=0.2\textwidth,align=c]{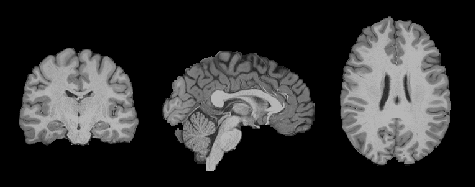}&
        \includegraphics[width=0.2\textwidth,align=c]{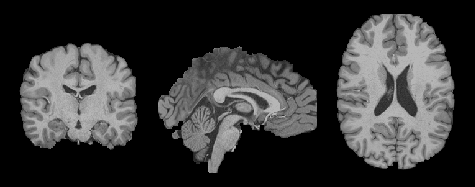}&
        \includegraphics[width=0.2\textwidth,align=c]{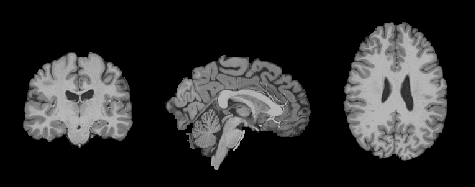}\\
    \bottomrule
    \end{tabular}    
    \caption{Example generated volumes.  Our slice-based model is able to generate realistic, if somewhat blurry, volumes at $256^3$, unlike the volumetric 3D $\alpha$WGAN model.}
    \label{fig:examples}
\end{figure}

We compare the generations produced by our model against a 3D VAE and other 3D generative network approaches from Kwon~\etal~\cite{kwon2019generation} at $64^3$, $128^3$, and $256^3$ sizes.  In models originally for $64^3$ inputs, we increase the number of layers to reach the desired output size. 

We use the following shorthand for describing architectures:
convolutional layer with N filters - \texttt{conv\_N},  
batch norm - \texttt{BN},
leaky ReLU - \texttt{LR},
max pooling - \texttt{MP},
reversible layer~\cite{gomez2017reversible} with 3 \texttt{conv\_16}  - \texttt{RL},
fully connected layer with N units - \texttt{FC\_N},
residual block with conv-ReLu-BatchNorm subblocks, halving size and doubling filters - \texttt{ResDown}. Compared models are:

\begin{description}
    \item[3D WGAN GP]~\cite{gulrajani2017improved}
    \item[3D VAE-GAN]~\cite{larsen2015autoencoding}
    \item[3D $\alpha$ GAN]~\cite{rosca2017variational}. For the model at $256^3$, we replace BatchNorm3D with InstanceNorm3D layers, and remove BatchNorm1D layers.
    \item[3D $\alpha$ WGAN] model proposed by Kwon~\etal~\cite{kwon2019generation}, with 1000 latent dimensions
    \item[3D VAE] our own implementation. Encoder and decoder are symmetric. Both mean and standard deviation have a fully connected layer. %
    \begin{description}
    \item[64$^3$ encoder] \texttt{Conv\_16 - BN - LR - MP - $3 \times $(RL - MP) - RL - FC\_512} 
    \item[128$^3$ encoder] \texttt{Conv\_16 - BN - LR - MP - $4 \times $(RL - MP) - FC\_1024} 
    \end{description}
    \item[Our proposed model]  We use a VAE with a 0.2 weight on the KL term, which produces better quality samples. Encoder and decoder are symmetric.
    \begin{description}
    \item[64$^3$ encoder] \texttt{Conv\_16 - BN - LR - $3 \times $(ResDown - LR) - ResDown} 
    \item[128$^3$ encoder] \texttt{Conv\_8 - BN - LR - $4 \times $(ResDown - LR) - ResDown} 
    \item[256$^3$ encoder] \texttt{Conv\_4 - BN - LR - $5 \times $(ResDown - LR) - ResDown}  
    \end{description}
    We used $N=400$ samples to estimate the sample means and covariances.
\end{description}

\subsection{Human Connectome Project Dataset}\label{sec:datasets}
We use T1w MR volumes from the Human Connectome Project (HCP)~\cite{van2013wu} dataset. To preprocess each brain,  we perform bias correction using the N4 algorithm~\cite{tustison2010n4itk} and  normalize the intensities per volume using the 1$^{st}$ and 99$^{th}$ percentiles (clipping the values at the lower and upper bounds). Skull stripping is performed by  FreeSurfer~\cite{fischl2012freesurfer}. We discard zero-filled planes to obtain a volume of size $256\times256\times256$ at $0.7$mm$\times0.7$mm $\times0.7$mm resolution, and bilinearly resample to the needed size.  We use coronal slices for training our method.
960 volumes are used for training and 40 for validation. 

\subsection{Training details}
We used the implementation from~\cite{kwon2019generation} for the baseline models evaluated in their paper.  For our proposed model, we used the Adam optimizer and performed a sweep of learning rates in $0.001$, 0.0001, 0.00001. We do not perform any augmentation during training.
To compute the RAS, we use a U-Net~\cite{ronneberger2015u} based segmentation network that was trained on 40 volumes of coronal brain slices using 15 labels. We used the Adam optimizer with default beta1, beta2, learning rate 0.001, and batch size 16, with a Dice training loss~\cite{milletari2016v}.

\begin{figure}[!htb]
    \centering
    \begin{tabular}{ccc}
    \includegraphics[width=0.3\textwidth]{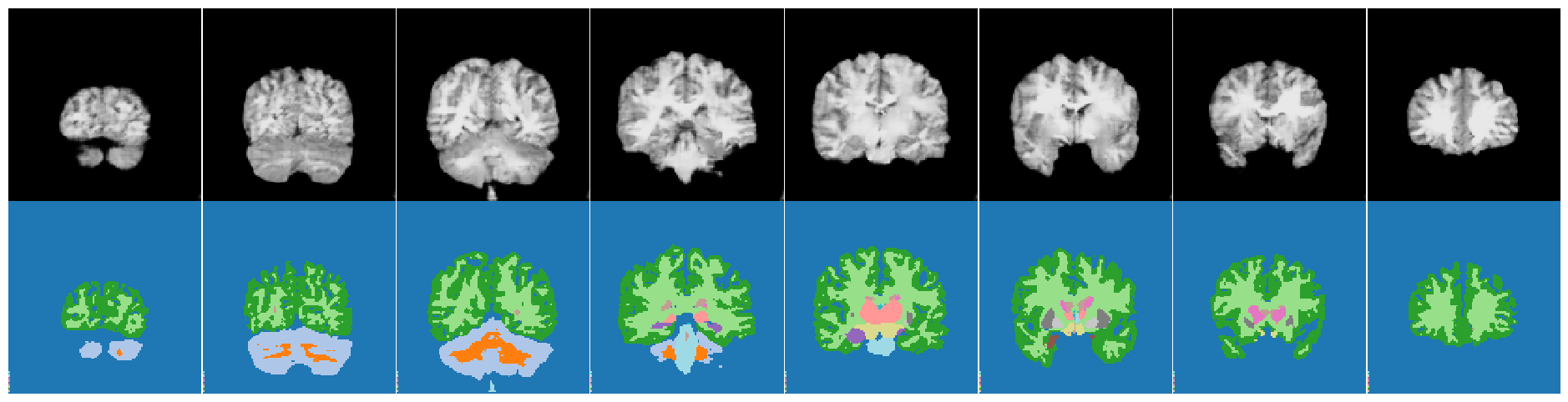} & 
    \includegraphics[width=0.3\textwidth]{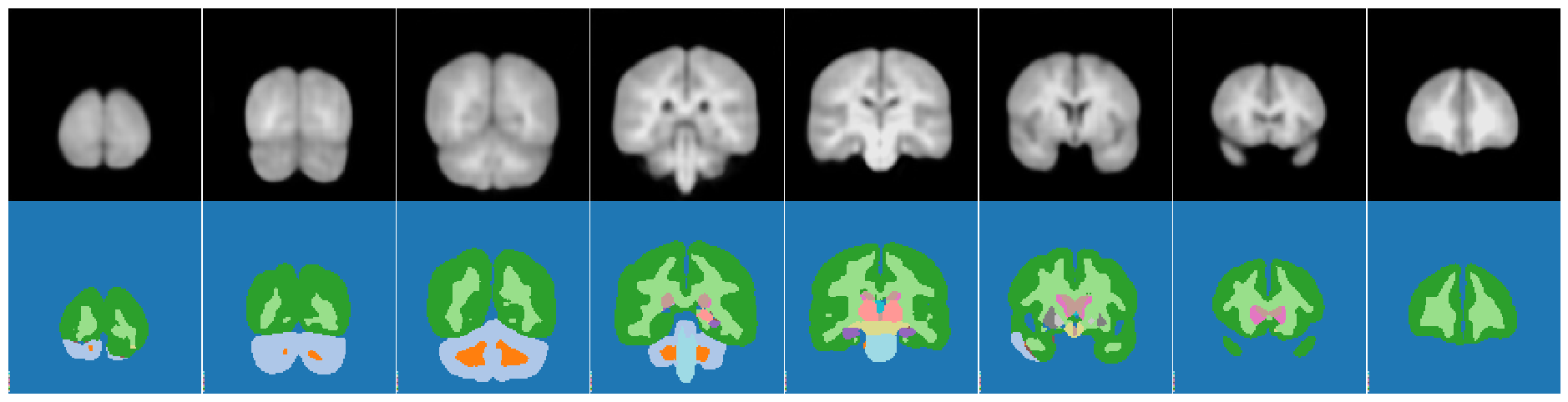} &
    \includegraphics[width=0.3\textwidth]{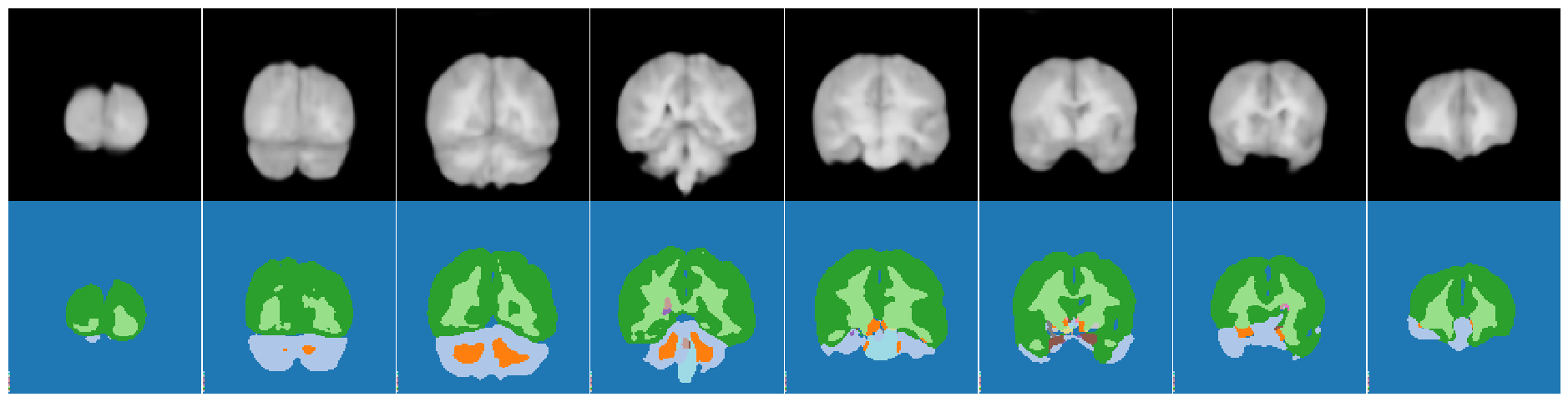}\\
    128$^3$ 3D $\alpha$WGAN & 128$^3$ Ours & 128$^3$ 3D VAE \\

    \includegraphics[width=0.3\textwidth]{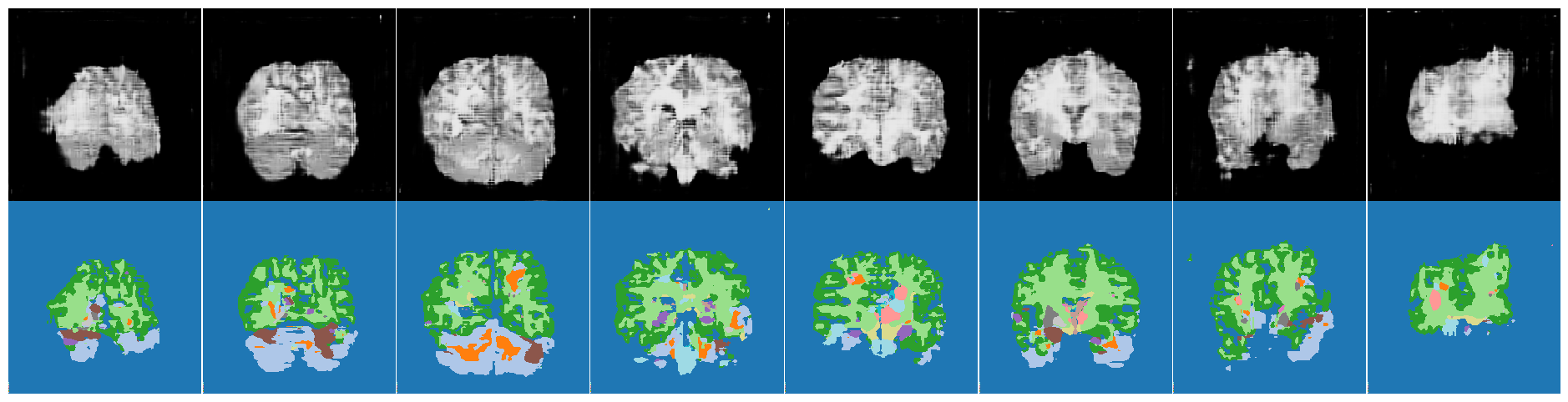}&    
    \includegraphics[width=0.3\textwidth]{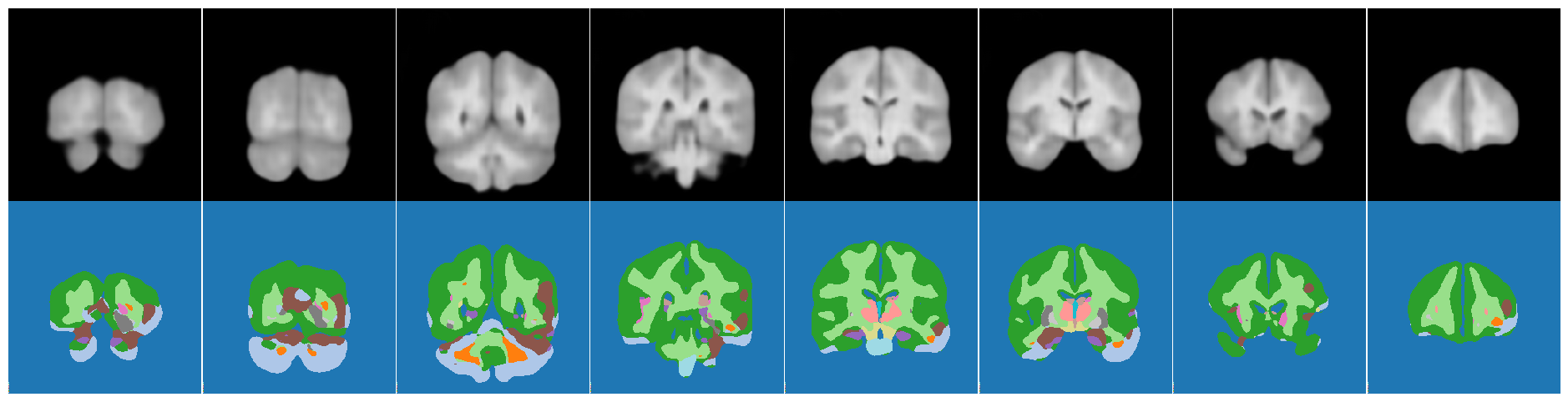} & 
    \includegraphics[width=0.3\textwidth]{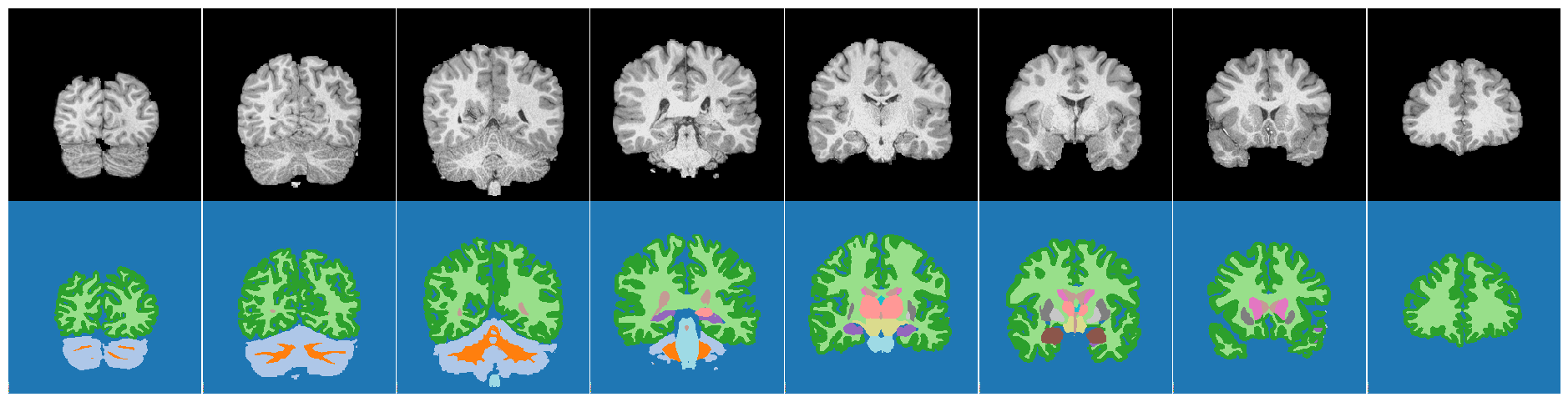} \\
    256$^3$ 3D $\alpha$WGAN & 256$^3$ Ours & 256$^3$ Real \\

    \end{tabular}
    \caption{Segmentations of example generated volumes. At 256$^3$ size, our model produces samples with more realistic segmentations than 3D $\alpha$WGAN.}
    \label{fig:segmentations}
\end{figure}

\section{Experimental Results}
\subsection{Example generations}

Figure~\ref{fig:examples} shows example generated volumes.  Our method is able to successfully sample consistent brain volumes.  Both our and the 3D VAE generated samples are somewhat blurry, which is a well-known shortcoming of VAE-based models.  We also see that our model can generate diverse brain shapes.  The 3D $\alpha$WGAN produces the visually highest quality samples at $128^3$, but fails to produce realistic samples at $256^3$, and suffers from blocky artefacts.

\begin{table}[]
\centering
\begin{tabular}{@{}lrrrrrrrrr@{}}
\toprule
 &
  \multicolumn{2}{c}{HCP 64$^3$} &
  \multicolumn{1}{l}{} &
  \multicolumn{2}{c}{HCP 128$^3$} &
  \multicolumn{1}{l}{} &
  \multicolumn{2}{c}{HCP 256$^3$} \\ \cmidrule(lr){2-3} \cmidrule(lr){5-6} \cmidrule(l){8-9} 
 &

  \multicolumn{1}{c}{MMD} &
  \multicolumn{1}{c}{MS-SSIM} &
  \multicolumn{1}{l}{} &
  \multicolumn{1}{c}{MMD} &
  \multicolumn{1}{c}{MS-SSIM} &
  \multicolumn{1}{l}{} &
  \multicolumn{1}{c}{MMD} &
  \multicolumn{1}{c}{MS-SSIM} \\
  \midrule
3D WGAN GP       & 14383 & 0.9995 &  &        &        &  &        &        & \\
3D VAE GAN       & 2054  & 0.9292 &  &        &        &  &        &        & \\
3D $\alpha$-GAN  & 7116  & 0.9848 &  &        &        &  &        &        & \\
3D $\alpha$-WGAN & 4488  & 0.8994 &  & 64446  &  0.9736&  & 912627 &  0.7106& \\
3D VAE           & 6823  & 0.9927 &  & 51476  & 0.9335 &  &        &        & \\
Ours             & 2396 & 0.9304  &  &  19890 &  0.9120&  & 323233 & 0.8768 & \\
Real             &       & 0.8786 &  &        &  0.7966&  &        & 0.7019 &\\ 
\bottomrule
\end{tabular}%

\caption{MMD and MS-SSIM for compared models. Our model produces samples close to the data distribution according to (low values of) MMD, and also generates diverse samples as measured by low MS-SSIM.}
\label{tab:metrics}
\end{table}

\subsection{Image diversity metrics}
We follow ~\cite{kwon2019generation} and report the Multiscale Structural Similarity (MS-SSIM) to measure the diversity of generated samples; and a minibatch estimate of Maximum Mean Discrepancy (MMD) to measure distance to the training distribution. We use the same settings as Kwon~\etal~\cite{kwon2019generation}. Due to computational cost, the MMD for $256^3$ was computed over 10 tests using batch size 4, instead of over 100 tests with batch size 8; and the $256^3$ MS-SSIM for real data is averaged over 5 tests, instead of over 20 tests. 
Table~\ref{tab:metrics} shows the MMD and MS-SSIM of the compared models. 

We compare all baseline models from~\cite{kwon2019generation} at $64^3$, and only the best-performing model from that set, the 3D $\alpha$WGAN, at larger sizes. Our 3D VAE at $256^3$ did not converge.

Our proposed method generates samples closest to the data distribution in the MMD sense at $128^3$ and $256^3$ sizes, and is second to 3D VAEGAN at $64^3$.  Our model also has a low MS-SSIM at $64^3$ and $128^3$, meaning the samples are diverse.  The MS-SSIM of the 3D $\alpha$WGAN at 256$^3$ is lower than ours because the MS-SSIM computes the pairwise similarity of generated samples only, and  the 256$^3$ 3D $\alpha$WGAN   generates very diverse but low-quality samples.

While MMD and MS-SSIM evaluate the samples in the distribution sense, they are not interpretable in terms of anatomical plausibility of the generated images. Thus the proposed RAS metric complements MMD and MS-SSIM.

\begin{figure}[!htb]
    \centering
    \includegraphics[width=1.0\textwidth]{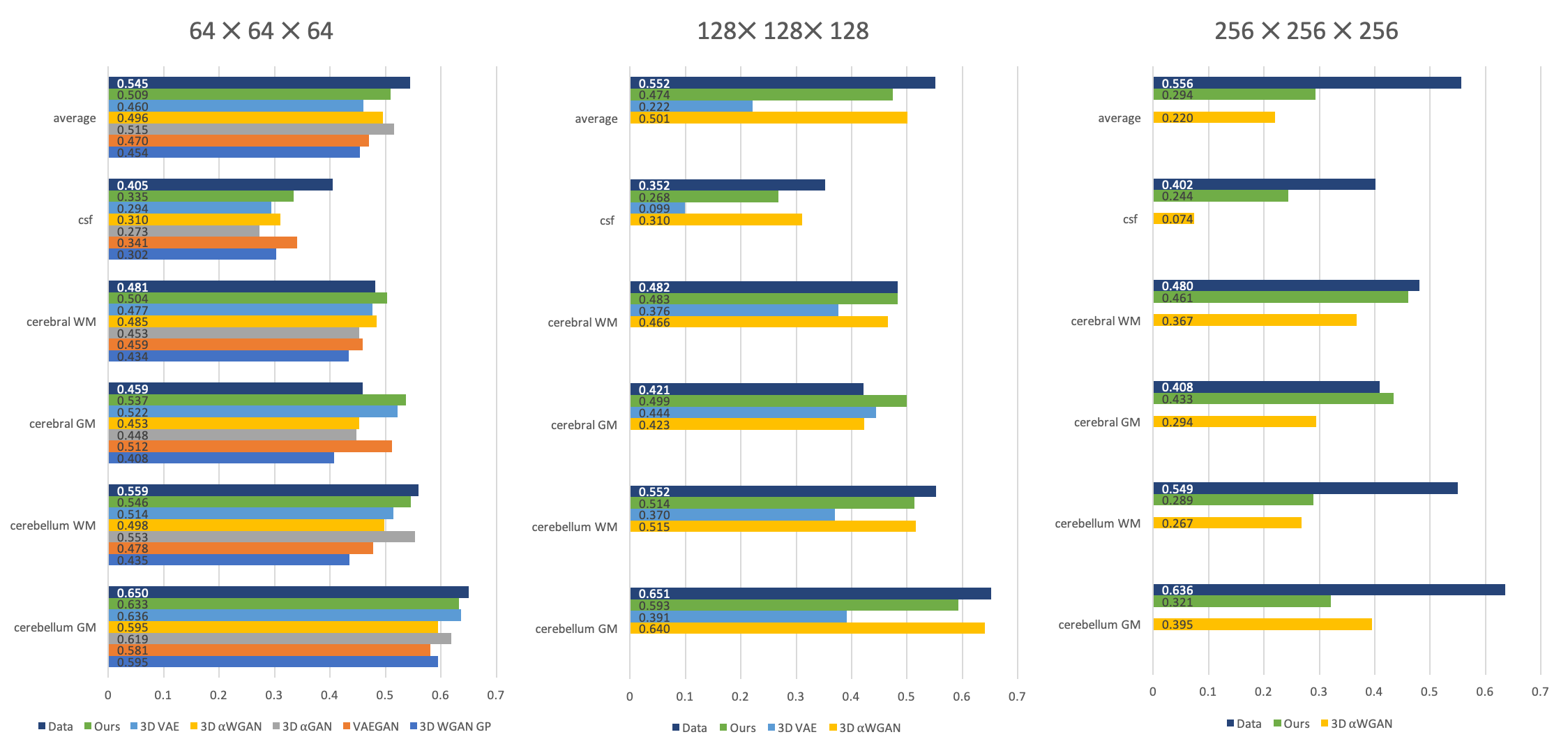}
    \caption{Realistic Atlas Score at different image sizes. Our model is competitive with other volumetric generation approaches at $64^3$ and $128^3$ sizes, and produces more realistic volumes than the 3D $\alpha$WGAN at $256^3$.}
    \label{fig:dice}
\end{figure}
\subsection{RAS evaluation}

Figure~\ref{fig:dice} shows RAS values. We also computed the RAS between different sets of real volumes to produce an upper bound. Both our model and 3D $\alpha$WGAN have similar performance at 128$^3$ size, while our model's samples are more realistic at 256$^3$ size. Figure~\ref{fig:segmentations} shows example segmentations from the compared models.

RAS values are affected by the quality of the inter-subject registration. For structures with high intersubject variability, the registration quality can be low for some pairs of real data, decreasing the average RAS. The synthetic examples often fail to create complex patterns in such structures, producing blurred areas, effectively simplifying the registration task and preventing RAS from dropping very low.  Notably this is a drawback of RAS and the reason why it should be considered as a complementary score to MMD and MSSIM. However, we note although RAS is insensitive to the diversity of generations, it effectively quantifies the realistic nature of generations in an interpretable manner.

\section{Discussion}

Taken together, the MMD, MS-SSIM and RAS evaluation show that the proposed model for approximating distributions of 3D volumes via 2D VAEs can produce realistic samples on par with or better than the state of the art GAN approaches, extending the capabilites of current VAE models. Our simple yet efficient approach opens up new avenues for building Bayesian models using 3D priors distributions, and provides a possible approach for modeling distributions at 256$^3$ image size.

\FloatBarrier

\bibliographystyle{splncs04}
\bibliography{egbib}

\begin{thebibliography}{10}
\providecommand{\url}[1]{\texttt{#1}}
\providecommand{\urlprefix}{URL }
\providecommand{\doi}[1]{https://doi.org/#1}

\bibitem{Ashburner2005}
Ashburner, J., Friston, K.J.: Unified segmentation. {NeuroImage}
  \textbf{26}(3),  839--851 (jul 2005). \doi{10.1016/j.neuroimage.2005.02.018}

\bibitem{chen2018unsupervised}
Chen, X., Konukoglu, E.: Unsupervised detection of lesions in brain mri using
  constrained adversarial auto-encoders. In: MIDL Conference book. MIDL (2018)

\bibitem{dice1945measures}
Dice, L.R.: Measures of the amount of ecologic association between species.
  Ecology  \textbf{26}(3),  297--302 (1945)

\bibitem{fischl2012freesurfer}
Fischl, B.: Freesurfer. Neuroimage  \textbf{62}(2),  774--781 (2012)

\bibitem{gomez2017reversible}
Gomez, A.N., Ren, M., Urtasun, R., Grosse, R.B.: The reversible residual
  network: Backpropagation without storing activations. In: Advances in neural
  information processing systems. pp. 2214--2224 (2017)

\bibitem{gulrajani2017improved}
Gulrajani, I., Ahmed, F., Arjovsky, M., Dumoulin, V., Courville, A.C.: Improved
  training of wasserstein gans. In: Advances in neural information processing
  systems. pp. 5767--5777 (2017)

\bibitem{heusel2017gans}
Heusel, M., Ramsauer, H., Unterthiner, T., Nessler, B., Hochreiter, S.: Gans
  trained by a two time-scale update rule converge to a local nash equilibrium.
  In: Advances in neural information processing systems. pp. 6626--6637 (2017)

\bibitem{KingmaW13}
Kingma, D.P., Welling, M.: Auto-encoding variational bayes. In: Bengio, Y.,
  LeCun, Y. (eds.) 2nd International Conference on Learning Representations,
  {ICLR} 2014, Banff, AB, Canada, April 14-16, 2014, Conference Track
  Proceedings (2014), \url{http://arxiv.org/abs/1312.6114}

\bibitem{kratzwald2017improving}
Kratzwald, B., Huang, Z., Paudel, D.P., Dinesh, A., Van~Gool, L.: Improving
  video generation for multi-functional applications. arXiv preprint
  arXiv:1711.11453  (2017)

\bibitem{kwon2019generation}
Kwon, G., Han, C., Kim, D.s.: Generation of 3d brain mri using auto-encoding
  generative adversarial networks. In: International Conference on Medical
  Image Computing and Computer-Assisted Intervention. pp. 118--126. Springer
  (2019)

\bibitem{larsen2015autoencoding}
Larsen, A.B.L., S{\o}nderby, S.K., Larochelle, H., Winther, O.: Autoencoding
  beyond pixels using a learned similarity metric. arXiv preprint
  arXiv:1512.09300  (2015)

\bibitem{Leemput2009}
Leemput, K.V., Bakkour, A., Benner, T., Wiggins, G., Wald, L.L., Augustinack,
  J., Dickerson, B.C., Golland, P., Fischl, B.: Automated segmentation of
  hippocampal subfields from ultra-high resolution in vivo {MRI}. Hippocampus
  \textbf{19}(6),  549--557 (jun 2009). \doi{10.1002/hipo.20615}

\bibitem{milletari2016v}
Milletari, F., Navab, N., Ahmadi, S.A.: V-net: Fully convolutional neural
  networks for volumetric medical image segmentation. In: 2016 Fourth
  International Conference on 3D Vision (3DV). pp. 565--571. IEEE (2016)

\bibitem{ronneberger2015u}
Ronneberger, O., Fischer, P., Brox, T.: U-net: Convolutional networks for
  biomedical image segmentation. In: International Conference on Medical image
  computing and computer-assisted intervention. pp. 234--241. Springer (2015)

\bibitem{rosca2017variational}
Rosca, M., Lakshminarayanan, B., Warde-Farley, D., Mohamed, S.: Variational
  approaches for auto-encoding generative adversarial networks. arXiv preprint
  arXiv:1706.04987  (2017)

\bibitem{salimans2016improved}
Salimans, T., Goodfellow, I., Zaremba, W., Cheung, V., Radford, A., Chen, X.:
  Improved techniques for training gans. In: Advances in neural information
  processing systems. pp. 2234--2242 (2016)

\bibitem{tezcan2018mr}
Tezcan, K.C., Baumgartner, C.F., Luechinger, R., Pruessmann, K.P., Konukoglu,
  E.: Mr image reconstruction using deep density priors. IEEE transactions on
  medical imaging  \textbf{38}(7),  1633--1642 (2018)

\bibitem{tustison2010n4itk}
Tustison, N.J., Avants, B.B., Cook, P.A., Zheng, Y., Egan, A., Yushkevich,
  P.A., Gee, J.C.: N4itk: improved n3 bias correction. IEEE transactions on
  medical imaging  \textbf{29}(6),  1310--1320 (2010)

\bibitem{Valindria2017}
Valindria, V.V., Lavdas, I., Bai, W., Kamnitsas, K., Aboagye, E.O., Rockall,
  A.G., Rueckert, D., Glocker, B.: Reverse classification accuracy: Predicting
  segmentation performance in the absence of ground truth. {IEEE} Transactions
  on Medical Imaging  \textbf{36}(8),  1597--1606 (aug 2017).
  \doi{10.1109/tmi.2017.2665165}

\bibitem{van2013wu}
Van~Essen, D.C., Smith, S.M., Barch, D.M., Behrens, T.E., Yacoub, E., Ugurbil,
  K., Consortium, W.M.H., et~al.: The wu-minn human connectome project: an
  overview. Neuroimage  \textbf{80},  62--79 (2013)

\bibitem{vondrick2016generating}
Vondrick, C., Pirsiavash, H., Torralba, A.: Generating videos with scene
  dynamics. In: Advances in neural information processing systems. pp. 613--621
  (2016)

\end{thebibliography}

\end{document}